# High Resolution Spectral Metrology Leveraging Topologically Enhanced Optical Activity in Fibers

Aaron P. Greenberg[1], Gautam Prabhakar[1], Siddharth Ramachandran[1]*

[1] Boston University, Boston, MA 02215, USA.

* Correspondence to: sidr@bu.edu

## Abstract

Optical rotation, a form of optical activity, is a phenomenon employed in various metrological applications and industries including chemical, food, and pharmaceutical. In naturally-occurring, as well as structured media, the integrated effect is, however, typically small. Here, we demonstrate that, by exploiting the inherent and stable spin-orbit interaction of orbital angular momentum fiber modes, giant, scalable optical activity can be obtained, and that we can use this effect to realize a new type of wavemeter by exploiting its optical rotary dispersion. The device we construct provides for an instantaneous wavelength-measurement technique with high resolving power $R = 3.4 \times 10^6$ (i.e. resolution < 0.3 pm at 1-μm wavelengths) and can also detect spectral bandwidths of known lineshapes with high sensitivity.

## INTRODUCTION:

It has been well established that light with helical phase fronts carry orbital angular momentum (OAM) as designated with $\pm\mathcal{L}$ topological charge[1,2], which defines the number of light's twists per wavelength and directional handedness. In recent years OAM beams have risen to prominence for their applications in STED microscopy[3], laser machining[4], advanced manufacturing[5], optical tweezers[6], and both classical and quantum communications[7-9]. These beams can also carry spin angular momentum (SAM), as determined by their circular polarization, where left or right is denoted by $\sigma^\pm = \hat{x} \pm i\hat{y}$, respectively. The two quantities, $\mathcal{L}$ and $\sigma$, and how they are paired together, define a 4-dimensional Hilbert space of given $|\mathcal{L}|$ in optical fibers [**Fig. 1(A)**], represented by the electric fields[10]:

$$E(r,\phi,z,t) = F(r) \begin{cases} \sigma^\pm \exp(\pm i\mathcal{L}\phi) \exp(i\beta_{SOa}z) \\ \sigma^\pm \exp(\mp i\mathcal{L}\phi) \exp(i\beta_{SOaa}z) \end{cases} e^{-i\omega t} \quad (1)$$

where $F(r)$ is the mode amplitude distribution, $\phi$ is the azimuthal angle, $\omega$ is the angular frequency, and $\beta$ is the propagation constant, which is proportional to the fiber mode effective index $(n_{eff})$ and wavelength $(\lambda)$ by $\beta = 2\pi \cdot n_{eff}/\lambda$. The distinct propagation constants, $\beta_{SOa}$ and $\beta_{SOaa}$, represent two degenerate spin-orbit aligned states (SOa) where $\mathcal{L}$ and $\sigma$ are of the same sign, and two degenerate spin-orbit anti-aligned states (SOaa) where these quantities are of opposite sign. The degeneracy within this Hilbert space is lifted between these SOa and SOaa pairs due to light confinement in a waveguide, and by the spin-orbit interaction (SOI) of light, which describes the interaction between OAM and SAM[11] in an attractive potential. The spatial inhomogeneity of the ring-core refractive index profile $n(r)$ of fibers designed for OAM mode propagation[12,13] [see Supplementary S1], introduces geometrodynamic polarization-dependent perturbations that splits effective index $(\Delta n_{eff})$ depending on $\sigma^{\pm}$:

$$\Delta n_{eff} \propto \mathcal{L} \int F^2(r) \frac{\partial \Delta n(r)}{\partial r} dr \quad (2)$$

where $\Delta n(r)$ is the vortex fiber's $n(r)$ profile, relative to the index of its cladding. The intrinsic OAM of the modes amplifies this effect, enabling large effective index differences $\Delta n_{eff} > 10^{-4}$ between $\sigma^+$ and $\sigma^-$ OAM modes with the same topological charge $\mathcal{L}$. Ergo, the SOI of light is responsible for splitting the effective indices of SOa and SOaa modes, resulting in *circular birefringence*[14] [**Fig. 1(A)**].

A superposition state of two modes with the same $\mathcal{L}$ but opposite $\sigma$ (i.e. a SOa and SOaa state) effectively yields a linearly polarized beam with topological charge $\mathcal{L}$. Hence, the SOI-induced circular birefringence results in *optical activity* (OA), which manifests as a rotation of the beam's linear polarization orientation angle as it propagates down the fiber [see Eq. 3 and **Fig. 1(B)**]. For an input with a $\hat{x}$ ($= \sigma^+ + \sigma^-$) polarized OAM superposition of topological charge $\mathcal{L}$, the fiber output field would be:

$$E(r, \phi, z, t) = F(r) \exp(-i\omega t) \exp(i\mathcal{L}\phi) \exp(i\bar{\beta}\mathbb{Z}) \{\sigma^+ e^{i\gamma} + \sigma^- e^{-i\gamma}\}$$

$$= F(r) \exp(-i\omega t) \exp(i\mathcal{L}\phi) \exp(i\bar{\beta}\mathbb{Z}) \{\hat{x} \cos\gamma - \hat{y} \sin\gamma\} \quad (3)$$

where $\mathbb{Z}$ is the fiber (interaction) length, $\bar{\beta}$ is an average propagation constant $\bar{\beta} = (\beta_{SOa} + \beta_{SOaa})/2$ and $\gamma$ is the output OA polarization angle:

$$\gamma = \mathbb{Z} \cdot (\beta_{SOa} - \beta_{SOaa})/2 = \mathbb{Z} \cdot \pi \cdot \Delta n_{eff}/\lambda \quad (4)$$

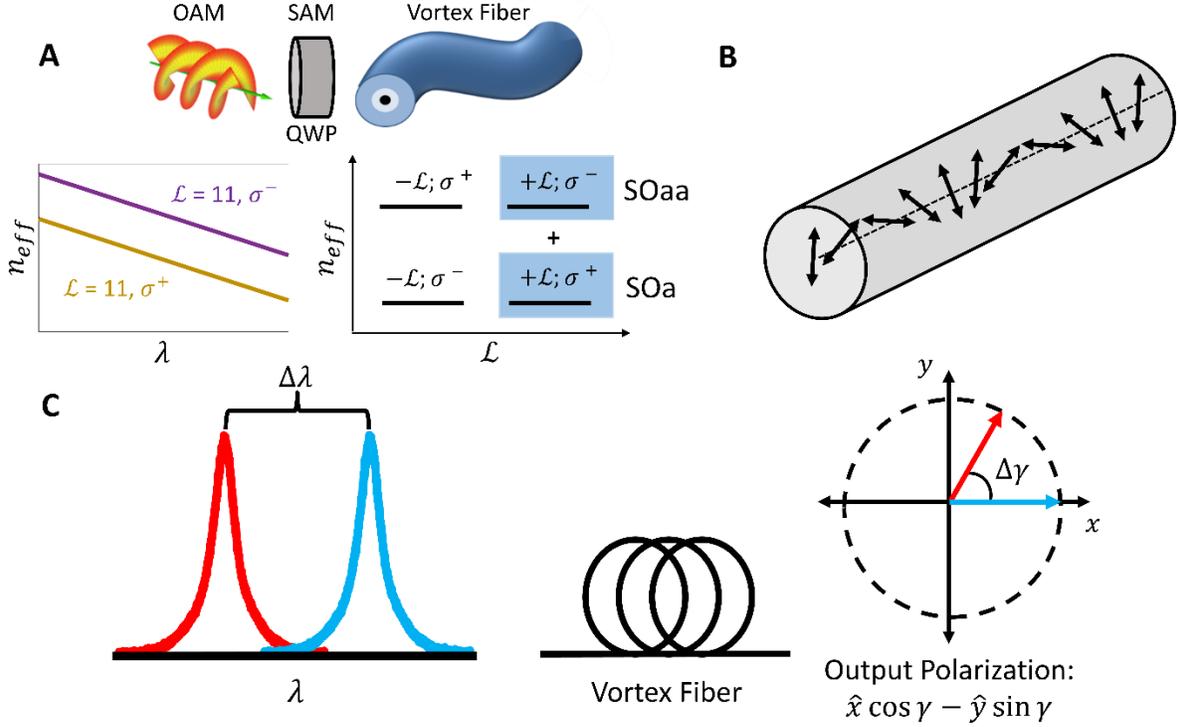

**Fig. 1 | Vortex fiber superposition phenomena. (A)** Spin orbit interaction (SOI) between OAM and SAM (imparted by quarter wave plate, QWP) splits effective index ($n_{eff}$) of SOa and SOaa states in a fixed-$|\mathcal{L}|$ 4-D vortex fiber mode Hilbert space. Numerical simulation reveals the circular birefringence between OAM $\mathcal{L} = +11$ modes with opposite $\sigma$. The superposition of the highlighted blue states is a linearly polarized OAM beam; which, when launched into the fiber, experiences *optical activity*. **(B)** Illustrative depiction of optical activity, where a linearly polarized beam rotates while propagating through the vortex fiber. **(C)** The wavelength dependence of optical activity, known as *optical rotary dispersion*, means a shift in wavelength ($\Delta\lambda$) rotates the output linear polarization ($\Delta\gamma$).

The polarization angle $\gamma$ exhibits wavelength dependence, known as *optical rotary dispersion* (ORD) [**Fig. 1(C)**], which has played roles in quantum metrology[15] and in ascertaining molecular conformation[16]. Although all OA media, either naturally occurring[17,18], or structured materials[19-22], share these traits, OAM vortex fibers provide for considerable scalability. Whereas the interaction length for these other cases is limited between $\mu m$ and $cm$, vortex fibers can support $km$-length propagation[23]. Another advantage is that $\Delta n_{eff}$ scales as $\mathcal{L}^2$ in these vortex fibers[24] [see Eq. 2], due to the combined effect of SOI (factor of $\mathcal{L}$) and the interaction of the optical field with the index inhomogeneity of the fiber $\left( \int F^2(r) \frac{\partial \Delta n(r)}{\partial r} dr \propto \mathcal{L} \right)$. Thus, it is apparent that by controlling the length of the fiber and choosing higher mode orders, we can achieve giant and scalable OA. In addition, by employing this OAM superposition state to obtain giant OA, we can observe fine changes in wavelength, making it the effect that forms the basis for our high resolution ORD Wavemeter.

**RESULTS:**

The inherent optical rotary dispersion of OAM superposition modes is the phenomenon used to establish a wavelength to polarization spectroscopic mapping. In order to see this, we take a first-order Taylor expansion of Eq. 4, yielding:

$$\gamma = \mathbb{Z} \cdot \pi \cdot \Delta n_{eff}/\lambda \rightarrow \Delta\gamma = -\left\{\mathbb{Z}\pi \frac{\Delta n_g}{\lambda^2}\right\}\Delta\lambda = \alpha\Delta\lambda \quad (5)$$

where $\Delta n_g$ is the difference in group index between SOa and SOaa modes, $\Delta\lambda$ is a change in wavelength, and $\alpha$ is a condensed factor defining the integrated birefringent strength of our ORD [complete derivation in Supplementary S2]. Evidently, any change in wavelength linearly maps to a change in polarization angle. Using OAM fiber modes, we can substantially increase the magnitude of $\alpha$, hence enhancing resolving power.

We use a tunable CW laser at 1028 nm with 100-kHz linewidth, which is transformed to the $\mathcal{L}$ = +11; $\sigma^+ + \sigma^-$ (or $\hat{x}$-polarized) OAM superposition state with a spatial light modulator (SLM) and coupled to a 40-m vortex fiber[13] [**Fig. 2(A)**]. Output mode spatial interferometry analysis[25] indicates >20-dB mode purity and >16-dB polarization extinction [details in Supplementary S3]. A polarization beam splitter (PBS) projects the output beam into two orthogonal polarization bins ($\hat{x}, \hat{y}$) and corresponding powers ($P_x, P_y$) are measured with photodetectors. The ratio of $P_x \propto \cos^2(\gamma)$ and $P_y \propto \sin^2(\gamma)$ [see Eq. 3] yields $\gamma$:

$$P_y/P_x = \tan^2(\gamma) \rightarrow \gamma = \tan^{-1}\left(\sqrt{P_y/P_x}\right) \quad (6)$$

The simplicity of Eq. 6 means $\gamma$ can be read-out *instantaneously*, requiring no post-processing, and allows a unique mapping of wavelength to polarization angle between an angular free-spectral-range (FSR) of 0 and 90°. For the $\mathcal{L}$ = 11 OA state, this is equivalent to a wavelength FSR of $\Delta\lambda = 46$ pm. To calibrate the device, we plot polarization rotation $\Delta\gamma = (\gamma - \gamma_0)$ versus a corresponding change in wavelength $\Delta\lambda = (\lambda - \lambda_0)$, where $\gamma_0$ and $\lambda_0$ are two initial reference states [see Eq. 5 and **Fig. 2(B)**]. The slope of a linear fit applied to this plot is our calibrated $\alpha$ factor, completing our $\lambda$ to $\gamma$ mapping (see Methods).

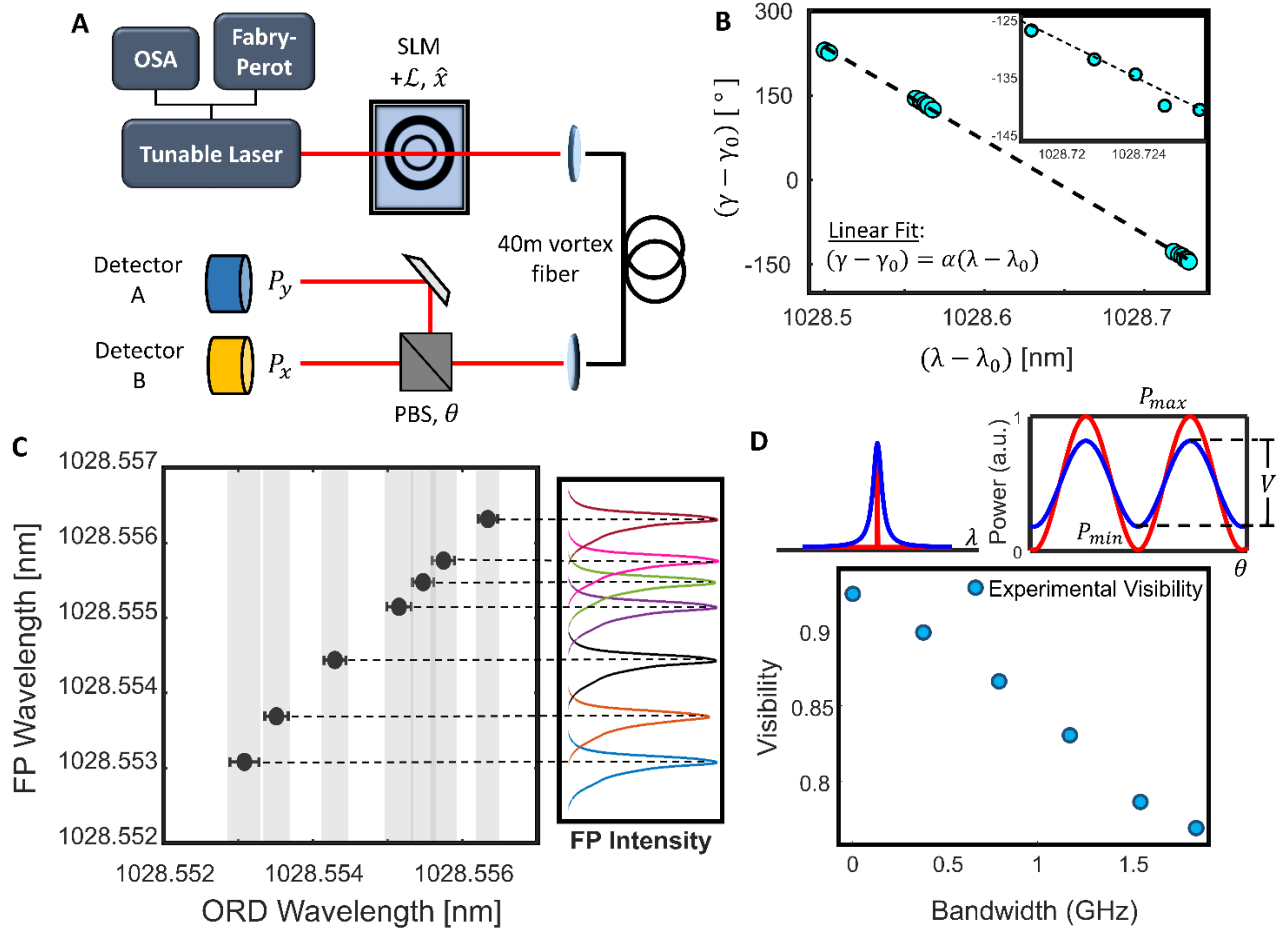

**Fig. 2 | ORD Wavemeter Performance. (A)** A SLM generated $\hat{x}$-polarized OAM superposition state experiences OA through 40 m of vortex fiber. At the output, a PBS splits power into two orthogonal polarization bins ($\hat{x}$, $\hat{y}$), and the ratio of powers, ($P_x$, $P_y$) instantaneously determines polarization orientation angle $\gamma$ (see Methods). **(B)** Calibration of ORD birefringent strength ($\alpha$) is necessary to accurately map $\lambda$ to $\gamma$. Wavelength sweep $\Delta\lambda$, resolved using a Fabry-Perot (FP) reference, has a corresponding polarization rotation $\Delta\gamma$, measured by power detectors. Data is collected over three spectral regimes (Inset focuses on one regime) and $\alpha$ is taken as the slope of an applied linear fit. **(C)** Resolution of the ORD wavemeter is determined by changing wavelength with a TEC in sub-picometer steps. The gray bands indicate detector dependent measurement error. Using OAM $\mathcal{L} = +11$ and 40-m vortex fiber we find the ORD wavemeter can measure down to $\Delta\lambda < 0.3$ pm at 1028 nm ($R = 3.4$ x $10^6$) with nearly full confidence thresholding. **(D)** ORD wavemeter visibility ($V$) reduces with broadening of laser linewidth. Red represents the visibility of a near-monochromatic laser, and blue represents a broadened spectrum. Measured visibility as a function of bandwidth reveals a monotonically decreasing relationship as the source is broadened by ~2 GHz.

To determine the resolving power of the ORD wavemeter, we finely tune our laser wavelength with a thermoelectric cooler (TEC) and measure $\gamma$. Using our calibrated $\alpha$ factor, we can convert the changes in polarization $\Delta\gamma$ to changes in wavelength $\Delta\lambda$, which we validate using a FP reference that simultaneously measures wavelength. The non-overlapping nature of the gray error bands in **Fig. 2(C)** confirm wavelength shifts down to at least 0.3 pm are readily discernable by our device. Since the measurement only involves recording power, the speed, noise performance, robustness, and reliability of the ORD wavemeter depends largely on the choice of photodetectors used (see Methods).

So far, we have shown the capability to measure wavelength to sub-picometer resolution, but the ORD wavemeter can detect spectral bandwidth as well[26]. By biasing the PBS angle $\theta$ in our setup to maximize and minimize powers in the two detectors ($P_{max}$, $P_{min}$), the bandwidth of the source may be measured from the visibility ($V = P_{max} - P_{min}/P_{max} + P_{min}$) of the ORD wavemeter. Linewidth broadening introduces additional spectral components that experience OA to differing degrees. The superposition of all these OA states turns the output into an elliptically polarized beam that reduces interferometric visibility [**Fig. 2(D)**]. With a 50-m vortex fiber we experimentally observe a clear monotonically decreasing relationship between bandwidth and visibility [**Fig. 2(D)**]. Considering we should be able to measure visibility down to $V \approx 0$, the small visibility change suggests total broadening detection much larger than the 2 GHz (~ 7 pm) we could measure with existing infrastructure. ORD wavemeter bandwidth detection could also be extended to measure larger spectral broadening by shortening fiber length and/or using lower order modes. This would change the power distribution of ($P_{max}$, $P_{min}$) and allow the ORD wavemeter to detect larger changes in bandwidth for the similar drops in visibility. In general, knowing the input functional form of the broadened source allows the measured visibility of the ORD wavemeter to be uniquely mapped to spectral bandwidth [simulations discussed in Supplementary S4].

We test the stability of the ORD wavemeter with respect to mechanical vibrations and temperature fluctuations. **Figure 3(A)** reveals that the measured reference rotation angle $\gamma_0$ (blue trace) remains invariant with time, even when subjected to mechanical vibrations. This is especially evident when compared with a similar experiment conducted with a conventional polarization–maintaining (PM) birefringent fiber of similar length, which can also be used to map polarization state to wavelength[27], but which displays *complete* instability (red trace). We also varied the temperature experienced by our device from 22 °C to 60 °C, and observe an oscillatory change in $\gamma_0$ [**Fig. 3(B)**]. This is primarily due to fiber expansion, which changes Z in Eq. 4, leading to periodic change in $\gamma_0$ [Eq. 6]. Hence, being predicable and systematic, ORD measurements with vortex fibers behaves dramatically differently from the chaotic behavior displayed by the PM fiber or, as we will discuss, other known interference-based systems.

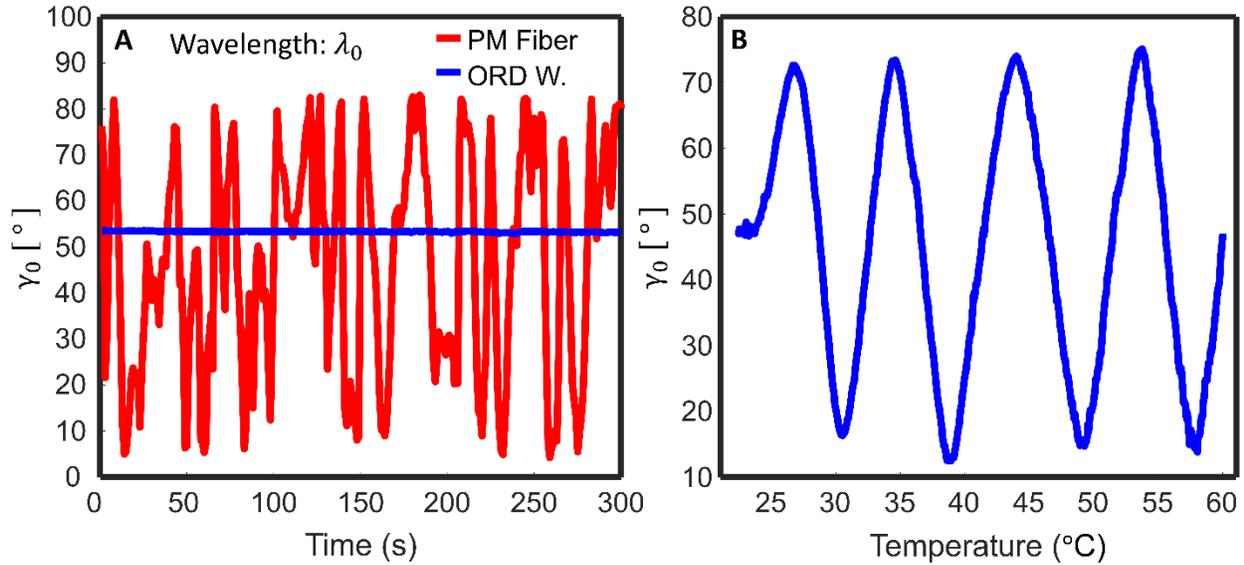

**Fig. 3 | Stability Metrics. (A)** Stability of the ORD wavemeter at a single wavelength ($\lambda_0$) in a mechanically perturbed environment, where the vortex fiber is shaken over a 5 minute period. The change in OA angle, $\gamma_0$, due to these perturbations is barely resolvable by our Fabry-Perot (< 0.4 pm). This is in contrast to the apparently unstable behavior of a birefringent PM-fiber wavemeter. **(B)** Stability of the ORD wavemeter at a single wavelength ($\lambda_0$) in a thermally perturbed environment, where the vortex fiber is heated from 22 °C to 60 °C. The thermal expansion of the fiber causes the OA angle to change with very systematic behavior. The mechanical and temperature stability tests both demonstrate how inherently resilient ORD wavemeter is to perturbed environments, unlike the PM-fiber which showed erratic instability.

## DISCUSSION:

The high resolving power of the demonstrated ORD wavemeter arises from the ability to obtain scalable OA in long lengths of vortex fibers, which, by definition, requires the stable propagation of superposition states. Singlet OAM states, as described by Eq. 1, are known to be stably propagated in vortex fibers[7,13]. But here, OA, provided by the $\sigma^{\pm}$ superposition of singlet OAM states, is essentially an interference phenomenon, which would usually be sensitive to environmental perturbations. For instance, even in *cm*-long conventional multimode fibers, modal coupling and speckle effects are well known. In fact, instability and loss of preservation of superposition states is a common debilitating feature of multiple systems [qubit-environment interactions[28], communications in turbulent atmospheres[29], speckle pattern spectrometers[30], etc.]. So the key question then arises – why is a superposition state in the vortex fiber stable under *perturbed* conditions, in contrast with expectations of other known interference phenomena? We surmise that the reason lies in the topological nature of vortex fiber birefringence. Whereas other means of yielding OA in potentially length-scalable fibers rely on material / strain effects or structured fabrication, the SOI-induced $\Delta n_{eff}$ splitting is *inherently* resistant to non-topological perturbations such as temperature, vibrations, and bends. This is to say, fiber perturbations would indeed affect the phase of the OAM eigenmodes[31], but such a phase change is identical for both SOa and SOaa modes that form the superposition yielding OA. Support for this reasoning is available by studying the form of SOI splitting [Eq. 2]. Topologically enhanced OA depends on the spatial overlap of the field (identical for SOa and SOaa modes) and the fiber's index profile *gradient*, $\partial \Delta n(r)/\partial r$, rather than the index profile $\Delta n(r)$ itself. The expected change that non-thermal perturbations have on the index gradient should be a second or higher order effect, and hence smaller. The experiments indicate that they are, in fact, negligible. The inherent stability of OA in the vortex fiber is key in obtaining scalability. In fact, this demonstration was limited to 40 m of fiber because our commercial FP interferometer could not provide higher resolution reference for comparison. Since OAM mode propagation up to 13.4 km[23] has been shown to be feasible in vortex fibers, we speculate that revolving powers $R$ two orders-of-magnitude higher than that demonstrated here are accessible by this device schematic.

It is instructive to compare the performance of the ORD wavemeter with existing alternative approaches. We do this in **Fig. 4(A)**, which shows resolving power of select representative spectroscopic devices distinguished in 4 distinct bands based on underlying technology. The first (red) band contains techniques generally comprising scanning / moving parts such as gratings[32], cavity mirrors[33], probes[34], and micro-electro mechanical systems[35] (MEMS). The second (orange) band contains devices that map wavelength to spatial position, then measured with a camera[36-40]. Devices that exploit the temporal response of a pulse[41-43], but whose speed would depend on specific pulse widths and shapes used, are shown in the third (blue) band. Finally, devices shown in the fourth (green) band perform the entire measurement using one, or a few, (single-pixel) detectors[44,45] – this category includes the use of PM fibers[27] ($R \sim 10^4$), as well as our ORD wavemeter ($R > 10^6$, that leverages fiber length scalability due to SOI). Devices in this green band are very promising for a range of high-speed monitoring and metrological applications, since single photodetectors can have bandwidths up to 100 GHz in the telecommunications window, yielding, potentially, ~ps response times.

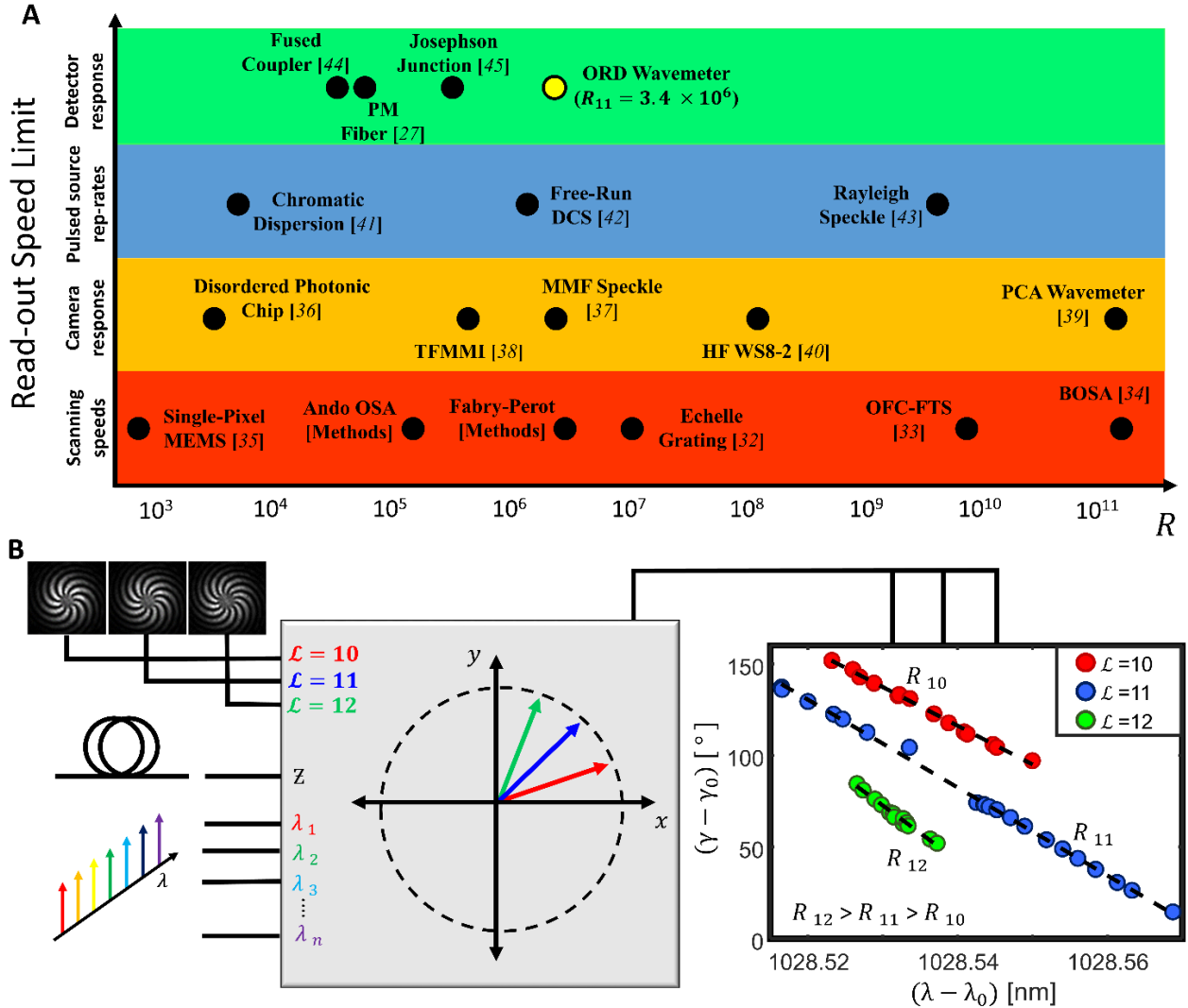

**Fig. 4 | (A)** Performance metric chart of various spectroscopic devices – wavemeters, spectrometers, interferometers – using resolving power and categorized speed constraints. Red band is for scanning devices, generally with mechanically moving parts. Orange is for devices that measure spatial intensity distributions with cameras. Devices whose speed depend on the pulse repetition rate of its laser source(s) are in the blue band. Lastly, green is for single-shot devices, which are principally speed constrained by the response time of their high speed photodetectors. The ORD wavemeter (indicated in gold) demonstrates the highest resolving power of a device belonging in the green band. **(B)** Multiplexing capabilities of the ORD wavemeter. Required inputs are OAM mode order $\mathcal{L}$ [Supplementary S7], vortex fiber length, and a wavelength sweep. For OAM of different $\mathcal{L}$ there are different strengths of optical rotation with wavelength, hence distinct $\alpha$ and resolutions ($R_{10}$, $R_{11}$, and $R_{12}$) can be identified. The multitude of available OAM modes in vortex fibers could be employed to develop generalized, multiplexed spectral measurement devices across wider wavelength ranges.

An additional benefit of the ORD wavemeter is that the vortex fiber can support multiple OAM modes of differing $\mathcal{L}$ simultaneously, making it ideal for multiplexing functionalities. ORD birefringent strength, $\alpha$, scales as $\mathcal{L}^2$ in these fibers[24] [see Eq. 2], as demonstrated in **Fig. 4(B)** [also see Supplementary S6], which shows distinct resolving powers ($R_{10}$, $R_{11}$, and $R_{12}$) for the different OAM mode orders in our fiber. With multiple detectors and well-known OAM mode sorting techniques[46,47] the schematic remains single-shot in nature. This is akin to deducing multiple unknowns given multiple measurands, as often deployed in multiparameter sensing. Given that vortex fibers supporting as many as 24 modes are now available[48], such multiplexing schematics could lead to additional functionalities, such as removing the FSR-to-resolution tradeoffs typical of interferometric devices, while remaining single-shot and high resolution.

## Methods

**ORD Wavemeter Construction and Experimental Setup**

**Figure S1** illustrates a fully detailed ORD wavemeter setup. Our experiments used a tunable, fiber-coupled, external cavity laser (ECL, Toptica DL-pro) with narrowband < 100 kHz linewidth. The laser is passed through an optical isolator, preventing back reflections, and a polarization controller (Polcon), allowing arbitrary control of polarization. A 3-dB coupler redirects the ECL to multiple paths, two of which (back-reflection and forward) couple to an optical spectrum analyzer (OSA, Ando AQ6317B) and a Fabry-Perot scanning interferometer (FP, Thorlabs SA210-8B) [further specifications in next section]. The other forward path projects the laser to free-space, which we convert to an OAM beam with topological charge $\mathcal{L}$ using a fork-pattern hologram generated on a spatial light modulator [SLM, Hamamatsu x10468-08][13]. Since the SLM is $\hat{x}$-polarization sensitive, we maximized the power of $\hat{x}$-polarized light using a combination of the Polcon and a polarization beam splitter (PBS). When our input $\hat{x}$-polarized OAM beam couples to our optically active vortex fiber, propagation rotates the linear polarization to a new orientation [see Eq. 3], as determined by the fiber length and OAM mode order. For our resolution experiments, we used a 40-m fiber and mode order of $\mathcal{L} = +11$. At the fiber output, a beam splitter (BS) sends part of the beam to a Thorlabs DCC1545M camera for modal purity analysis[25] [see Supplementary S3 for details]. An output PBS divides the beam into two orthogonal polarization bins aligned with $(\hat{x}, \hat{y})$. In the $\hat{x}$-bin, we send the beam to an output SLM and iris aperture to perform mode conversion [see Supplementary S3 for details]. The $\hat{y}$-bin also undergoes mode conversion, but with the addition of a half wave plate (HWP) to change polarization from $\hat{y}$ to $\hat{x}$. Following mode conversion, two low noise Ge photodetectors (Agilent HP 81521B) measure each polarization bin's respective powers $P_x$ and $P_y$, which are used to calculate the OA polarization angle $\gamma$ [see Eq. 6] for our ORD wavemeter measurements. Slight setup variations are used for spectral broadening and stability experiments, as discussed in Supplementary sections S4 and S5.

**Calibration Methodology, Mapping Accuracy, and Noise Characterization**

To calibrate our device and complete our spectroscopic mapping, we simply need to determine our ORD birefringent strength, $\alpha$, such that the linear relationship $\Delta\gamma = \alpha\Delta\lambda$ is satisfied [see Eq. 5]. For our calibration, we measure rotations in polarization $\Delta\gamma = (\gamma - \gamma_0)$ corresponding to changes in wavelength $\Delta\lambda = (\lambda - \lambda_0)$ over three mode-hop free spectral regimes of the ECL [see **Fig. 2(B)**]. The reference, or initial, wavelength, $\lambda_0$ was measured using a commercial high-resolution OSA, which provided NIST-traceable absolute wavelength measurements with 10-pm resolution. Accordingly, the reference angle $\gamma_0$ is the initial polarization state of the output OA beam at $\lambda_0$. Subsequent relative wavelength measurements $\lambda$ were obtained with a commercial FP with ~ 0.25-pm resolution, and polarization angles $\gamma$ were measured from Eq. 6 with photodetectors. Since Equation 5 limits the angular FSR of the ORD wavemeter (using a single mode) to a polarization mapping between 0 to 90° (equivalent to a wavelength FSR $\Delta\lambda$ = 46 pm with the $\mathcal{L} = 11$ OA mode), the $\gamma$ measurements in each regime are unwrapped (with further wavelength validation from both the Fabry-Perot and OSA) past these limits, such that they form the linear relation as described in Eq. 5. The calibration factor $\alpha$ is simply the slope of a linear fit applied to the measurements described, and once established, can be used to accurately convert $\gamma$ to $\lambda$. When applying the linear fit, outliers arising from in-fiber mode coupling and detector noise floors were not included. Since the ORD wavemeter demonstrates considerable stability, as demonstrated in **Fig. 3** and discussed in Supplementary S5, once the device is calibrated we can expect that $\alpha$ will remain unchanged and reliable, even in perturbed environments.

After calibration, we require another test to determine the accuracy of the ORD spectroscopic mapping. At two mode-hop free spectral regimes, we measure wavelength deduced from the ORD wavemeter using our calibration factor and plot versus the presumed ground truth as measured by the FP [see **Fig. S2**]. Root-mean-squared error analysis around the ground truth slope = 1 line reveals that our calibrated device is capable of predicting wavelength with an accuracy of approximately 1 pm. As such, the ORD wavemeter is capable of high accuracy in determining wavelength after being appropriately calibrated.

For the Agilent HP 81521B Ge detectors used in this work, the estimated noise equivalent power was $< 50$ pW (with averaging time of 1s) when the ORD wavemeter resolution was measured to be 0.3 pm at $\lambda = 1028$ nm. However, other sources of noise exist, arising for instance, from PBS extinction ratios, SLM mode conversion efficiency, etc. Instead of independently characterizing each noise source, we provide an overall systems noise with the measurement-uncertainty band in Fig. 2(C) deduced from the statistics of repeated measurements.

**Acknowledgements**

The authors thank P. Kristensen for fabricating the vortex fiber used in these experiments, and Zelin Ma and Xiao Liu for insightful discussions. This work was supported by the Vannevar Bush Faculty Fellowship (N00014-19-1-2632), the Brookhaven National Labs (Contract: 354281), the National Science Foundation (ECCS-1610190), and the Air Force Office of Scientific Research (FA9550-14-1-0165).

**Author contributions**

GP and SR conceptualized the experiments, APG, GP, and SR designed, and APG and GP performed the experiments. APG and SR analyzed the data. APG, GP and SR contributed to manuscript preparation.

**Competing interests**

The authors declare no competing interests

**Data Availability**

Supplementary Information is available for this paper. Correspondence and requests for materials should be addressed to S. Ramachandran. All data related to the experiments described in this manuscript are recorded in laboratory notebooks of members of S. Ramachandran's group, and all associated digital data are stored on networked computers at Boston University, whose contents are archived periodically. This data is available upon request.

# Supplementary Materials

**This PDF file includes:**
Supplementary Method Figures
Supplementary Text
Figs. S1 to S7

S1 discusses the refractive index profile and properties of our optical fiber, known as vortex fiber. The full first-order Taylor expansion that generates Equation 5 is derived in S2, showing the ORD mapping. In S3 we explain how we performed modal purity measurements and eliminated degenerate mode coupling noise in our system. The modified visibility experimental setup, as well as numeric simulations for visibility, are discussed in S4. The ORD stability test methodology is described in S5, which covers mechanical and thermal perturbations. Experimental validation of the spin-orbit interaction's (SOI) $\Delta n_g$ modal dependence is shown in S6. In S7, we describe how we produced the spiral images used in **Fig. 4(B)**.

# Supplementary Method Figures

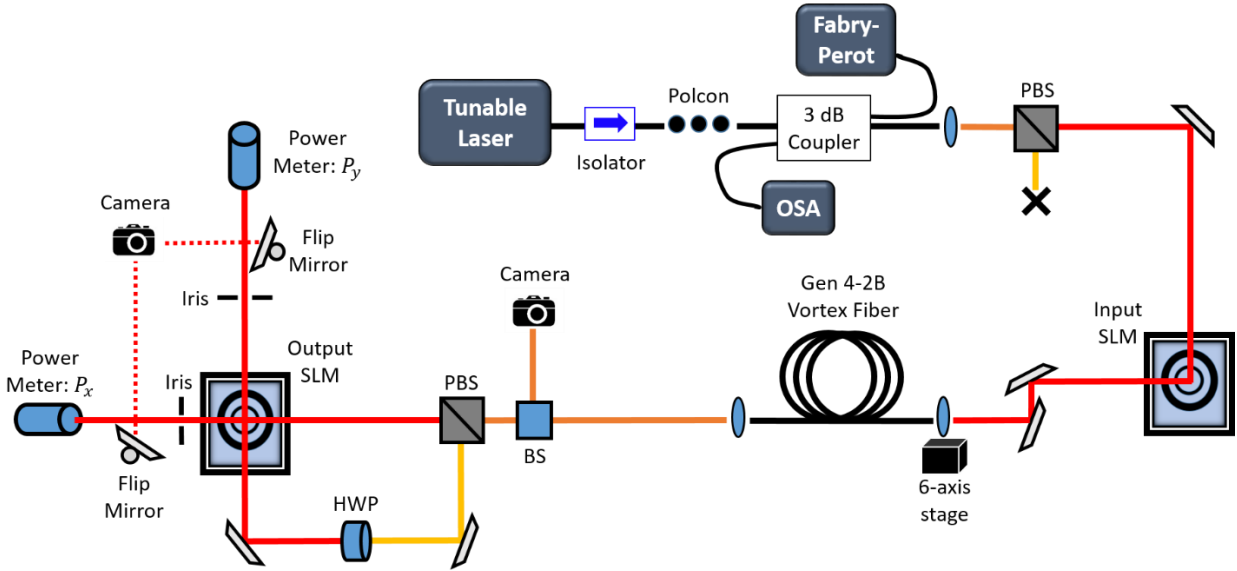

**Fig. S1.** Detailed ORD wavemeter resolving power experiment, expanding on the simplified version portrayed in Fig. 2(A). Free-space light polarization is indicated with color: Red is $\hat{x}$-polarization, yellow is $\hat{y}$-polarization, and orange is a combination of $\hat{x}$- and $\hat{y}$-polarization. OSA – optical spectral analyzer; PBS – polarization beam splitter; Polcon – polarization controller; SLM – spatial light modulator; BS – beam splitter; HWP – half wave plate.

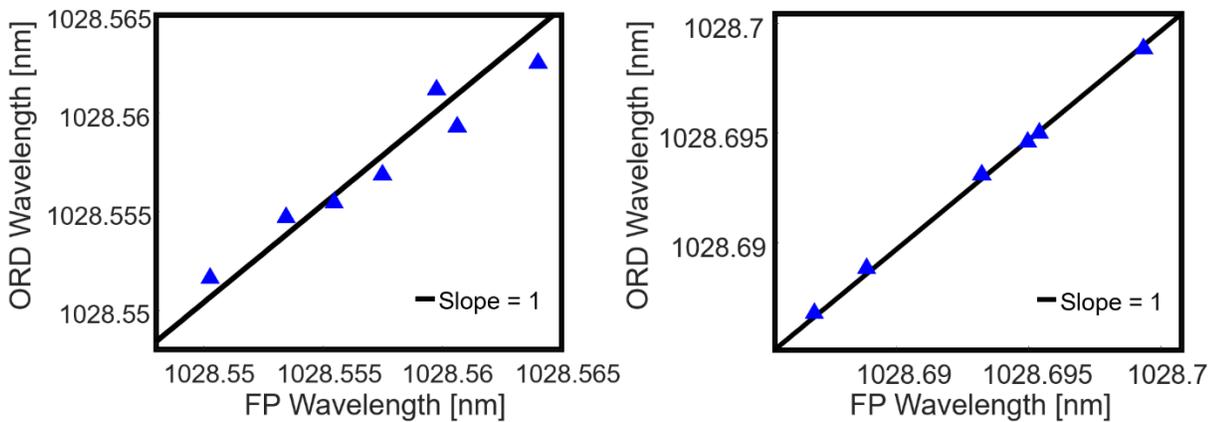

**Fig. S2.** Two mode-hop-free spectral regimes used to calculate accuracy of our ORD $\gamma$ to $\lambda$ mapping. Root-squared-mean error (RMSE) is used to determine the deviation of our ORD wavemeter measurements from the Fabry-Perot ground truth slope=1 line.

## Supplementary Text

### 1) Vortex Fiber Index Profile

Cylindrically symmetric and isotropic optical fibers utilizing a ring-core design fall under a particular classification, known as vortex fibers. Specifically, these fibers are specially designed to mirror the field profile of an OAM beam, which has been found to maximize the effective index splitting ($\Delta n_{eff}$) between modes of a given $|\mathcal{L}|$ but opposite circular polarizations, $\sigma^{\pm}$ (i.e. SOa and SOaa)[10]. We illustrate this profile in **Fig. S3**, which has been measured using an interferometry-based fiber profiler (Interfiber Analysis IFA-100) at 633nm. This split is due to spin-orbit coupling and the inhomogeneity of the ring-core, which, as predicted by perturbative theory, introduces polarization-dependent perturbations based on $\sigma^{\pm}$ and is enhanced by OAM[11,12]. As such, a 4-dimensional Hilbert space of OAM / SAM states exist in vortex fiber [see **Fig. 1(A)**] comprised of degenerate pairs of SOa and SOaa modes, which have effective index differences of $\Delta n_{eff} > 10^{-4}$ (leading to our observed circular birefringence[14]). A combination of this spin-orbit interaction $\Delta n_{eff}$ splitting and conservation of angular momentum allows vortex fibers to avoid both non-degenerate and degenerate coupling effects[13,31] between all the modes in their Hilbert space, and have been well known to stably propagate OAM / SAM singlet states[7], as described in Eq. 1.

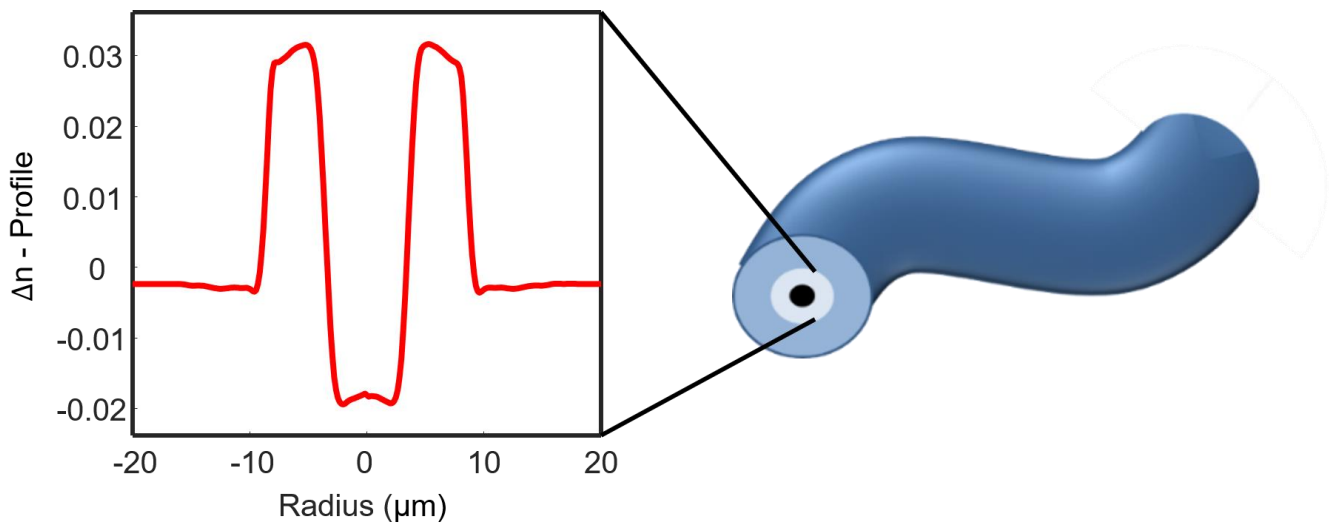

**Fig. S3. Vortex Fiber Profile.** The vortex fiber refractive index profile (red trace) is designed in the shape of an annular ring, mirroring the intensity profile of an OAM beam. Since the vortex fiber is cylindrically symmetric, the profile can be measured radially in 1-dimension. The $\Delta n$ refers to the profile's refractive index difference to standard silica.

2)  Wavelength to Polarization Mapping

The basis of the ORD wavemeter is the ability to map changes in wavelength directly to a rotation in polarization. The basic optical activity (OA) relation, for a single wavelength, is described in Eq. 4, and explicitly reveals a wavelength-dependence. However, to track how our OA polarization state changes with wavelength from a tunable laser source, we must also account for that fact that our effective index splitting $(\Delta n_{eff})$ is wavelength dependent. As such, we take a simple first order Taylor expansion of $\gamma(\lambda)$ [see Eq. 4] around $\lambda_1$:

$$\gamma(\lambda) = \frac{\text{Ƶ}\pi\Delta n_{eff}}{\lambda} \rightarrow \Delta\gamma = \left.\gamma(\lambda)\right|_{\lambda_1} - \gamma(\lambda_1) = \frac{\text{Ƶ}\pi\Delta n_{eff}}{\lambda_1} + \frac{1}{1!}\frac{d}{d\lambda}\left(\frac{\text{Ƶ}\pi\Delta n_{eff}}{\lambda}\right)(\lambda - \lambda_1) - \frac{\text{Ƶ}\pi\Delta n_{eff}}{\lambda_1}$$

$$= \frac{d}{d\lambda}\left(\frac{\text{Ƶ}\pi\Delta n_{eff}}{\lambda}\right)\Delta\lambda = \text{Ƶ}\pi\Delta\lambda\left(\frac{1}{\lambda}\frac{d\Delta n_{eff}}{d\lambda} - \frac{\Delta n_{eff}}{\lambda^2}\right) = -\frac{\text{Ƶ}\pi\Delta\lambda}{\lambda^2}\left(\Delta n_{eff} - \lambda\frac{d\Delta n_{eff}}{d\lambda}\right)$$

$$= -\frac{\text{Ƶ}\pi\Delta n_g}{\lambda^2}\Delta\lambda = \alpha\Delta\lambda \quad \text{(S1)}$$

We see that from the first order Taylor expansion of our OA, Eq. S1, we obtain the linear ORD mapping between $\Delta\gamma$ and $\Delta\lambda$ from Eq. 5.

3)  Modal Purity Analysis and Output Mode Conversion

The ORD wavemeter performance relies on our ability to propagate pure OAM states in vortex fiber[13]. To determine modal purity, the output OAM beam's intensity profile is imaged on a Thorlabs DCC1545M camera [see **Fig. S1**] and interferometric analysis is used to detect interference effects between vortex fiber modes[25]. One way to reduce modal cross talk is with precise input coupling of the beam to the vortex fiber. For this reason, a 6-axis stage (Thorlabs, MAX607L) and "walk-the-beam" method are used for optical alignment [see **Fig. S1**]. With proper alignment we achieved >20-dB mode purity between OAM modes of different topological charges $\mathcal{L}$. However, our analysis also revealed some degenerate coupling caused by fiber perturbations. This degenerate coupling transitioned OAM states $(+\mathcal{L}, \sigma^{\pm})$ to $(-\mathcal{L}, \sigma^{\mp})$, resulting in a ~10-dB polarization extinction ratio (PER) for the output beam.

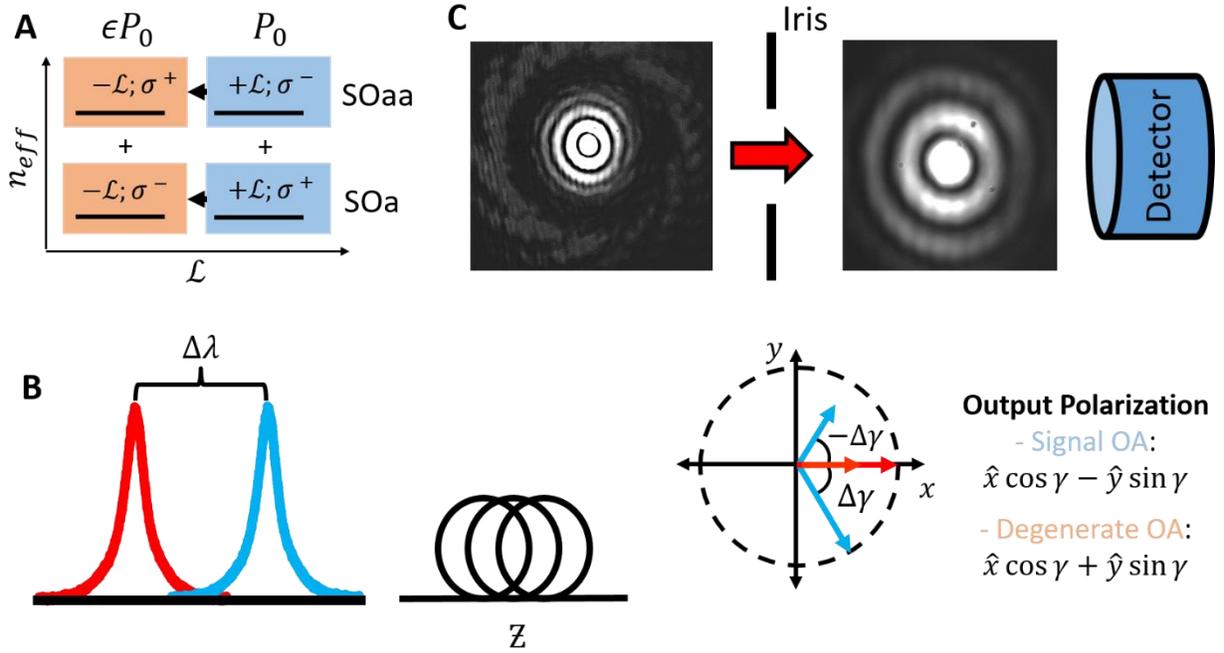

**Fig. S4. Degenerate coupling effects. (A)** Degenerate coupling fills the 4-D vortex fiber mode Hilbert space. The superposition of the highlighted orange states generate a second linearly polarized OAM mode, which undergoes degenerate optical activity (DOA) in vortex fiber **(B)** The optical rotary dispersion properties of OA and DOA. For a change in wavelength ($\Delta\lambda$), the OA polarization rotates $\Delta\gamma$, whereas the DOA polarization counter-rotates $-\Delta\gamma$. **(C)** A mode converted beam consisting of an OA Bessel beam and DOA $-2\mathcal{L}$ halo. An iris aperture blocks the $-2\mathcal{L}$ halo, ensuring that only light from the Bessel beam, and thus our original OA state, is measured by the photodetector.

Assuming equal fractional power, $\epsilon P_0$, degenerately couples into $-\mathcal{L}$-OAM SOa and SOaa states [see **Fig. S4(A)**], then another linearly polarized superposition state is generated which undergoes optical activity in vortex fiber. This degenerate optical activity (DOA) has identical ORD birefringent strength, since polarization changes at the same rate [see Eq. 5], but rotates with opposite handedness as compared to the original OA state [see **Fig. S4(B)**]. The counter-rotation of the OA and DOA alters the projection of power of the output beam into $(\hat{x}, \hat{y})$ polarization bins. This changes the OA $\gamma$ value calculated from Eq. 6, thus violating the mapping of our ORD wavemeter. To rectify this issue, we placed an SLM with a $-\mathcal{L}$ fork-pattern at our output [see **Fig. S1**] to perform a mode conversion technique that eliminates DOA contributions. This technique works by inducing a $-\mathcal{L}$ topological charge on our output OAM beam, transforming the spatial distributions of its OA and DOA components. The OA state's $\mathcal{L}$ topological charge is negated and it is converted into a $\mathcal{L} = 0$ Bessel beam in the far field[49]. Meanwhile, the DOA state's $-\mathcal{L}$ topological charge is doubled, converting it into a $-2\mathcal{L}$ "halo" with a large optical vortex. Adding iris apertures, we block the $-2\mathcal{L}$ halos surrounding the Bessel beams, allowing only light from the OA states to pass through to our photodetectors [see **Fig. S4(C)**]. Flip mirrors and a camera [see **Fig. S1**] were used to check that we properly eliminated these DOA halos. In principle, this mode conversion technique is not just limited to filtering out degenerate modes, but could filter out any undesired mode contributions as well (including non-degenerate OAM modes with different and unwanted topological charges). Using mode conversion, we achieved >16-dB PER and eliminated the effects of degenerate coupling on the ORD wavemeter performance. We note that several ring-core OAM fibers naturally provide >>16-dB PER, and hence this additional mode filtering is not strictly necessary for our technique to work. Rather, it serves to improve the ORD wavemeter's sensitivity.

### 4) Spectral Broadening Experiment and Simulations

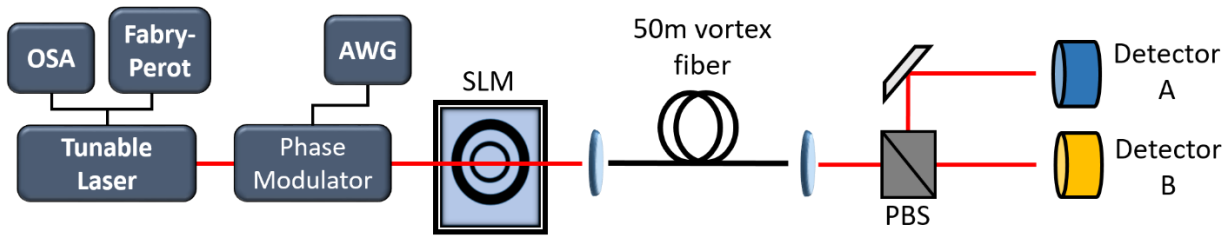

**Fig. S5.** Simplified depiction of the spectral broadening experiment with the ORD wavemeter. It is identical to the resolving power experiment except that fiber length has been changed to 50 m, and an arbitrary waveform generator (AWG) and electro-optic phase modulator have been added to control the broadening of our spectral lineshape.

The ORD wavemeter setup for measuring spectral broadening is the same as described in the Methods section, with a few alterations[26]. In order to control the linewidth of our laser, we add an electro-optic phase modulator (iXblue MPX-LN-10) controlled by an arbitrary waveform generator (AWG, Tektronix AWG7082C). **Figure S5** depicts a simplified setup, similar to **Fig. 2(A)**, to illustrate where these additional components are inserted. The spectral broadening experiments also requires our PBS to be biased at an angle $\theta$ such that power is maximized ($P_{max}$) in one polarization bin, while minimized ($P_{min}$) in the other. In our case, we use an alternative method and left $\theta = 0°$, aligned with ($\hat{x}, \hat{y}$) polarizations, and simply adjusted the operating wavelength until our power condition was satisfied. With either method, pre-processing of our setup to meet the visibility power condition is necessary, but could be accounted for during a calibration of the device.

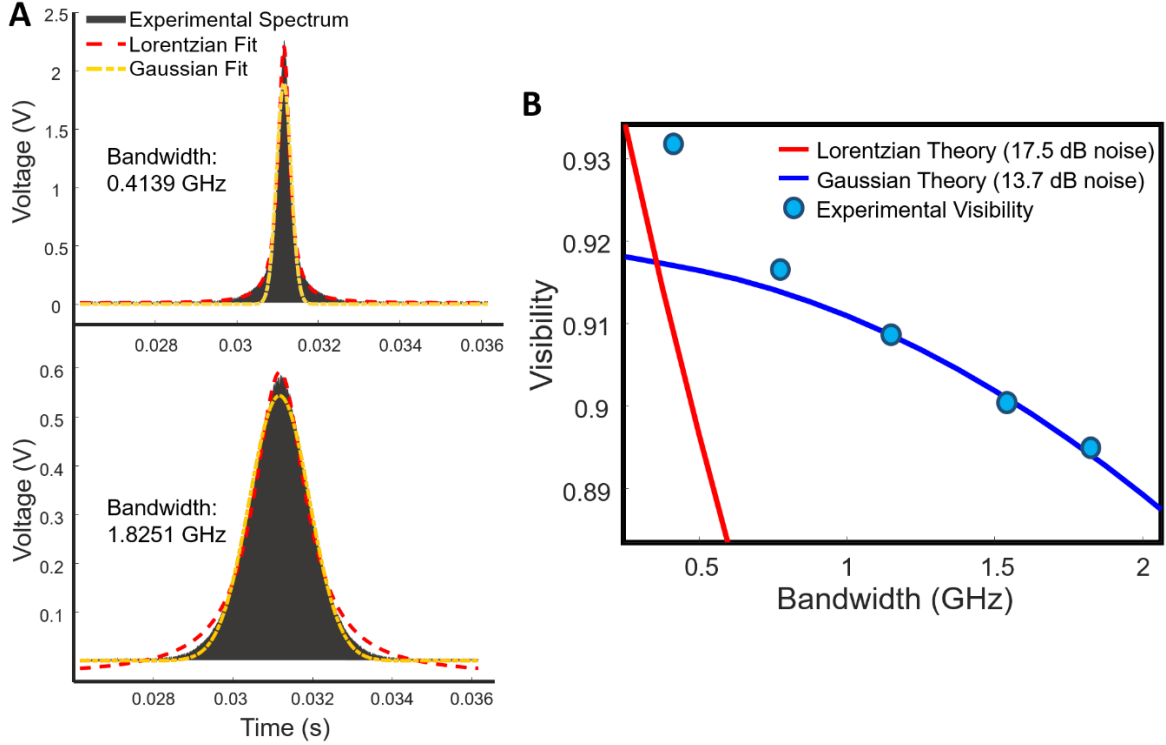

**Fig. S6. (A)** Lorentzian (red) and Gaussian (yellow) fits applied to both an unbroadened (bandwidth <0.7 GHz) and broadened (bandwidth >0.7 GHz) spectrum, measured with a FP. In the unbroadened case, a Lorentzian functional form fits the experimental spectrum better. In the broadened case, a Gaussian functional form fits the experimental spectrum better, especially around the tails. **(B)** Visibility simulations using both Lorentzian and Gaussian numeric models, considering different noise parameters. When the spectrum bandwidth is <0.7 GHz, the experimental visibility displays Lorentzian behavior, but as the spectrum is broadened, tends towards Gaussian behavior, albeit with a higher noise floor.

If we know the functional form of the ECL, then we can numerically model how visibility reduces with spectral bandwidth FWHM ($\delta\lambda$). Although the unbroadened ECL spectrum starts as a Lorentzian functional form, this is not maintained when the beam undergoes spectral broadening. This is due to the discrete nature of the AWG, which is resolution-limited (125 ps) in controlling phase modulation that generates an ideal broadened Lorentzian (log-Cauchy distribution). As such, the lineshape appears to broaden into a Lorentzian-Gaussian hybrid [**Fig. S6(A)**], with the function's peak fitting with a Lorentzian but tails trending towards Gaussian behavior. For sufficiently broadened spectra we can assume the functional form of our ECL is actually Gaussian, and therefore we can calculate $P_{max}$ to be:

$$P_{max} = P_0 \int_{\lambda} e^{-(\lambda-\lambda_0)^2/(\delta\lambda^2/4\ln(2))} \cos^2\left[ \mathsf{z}\pi\Delta n_g \left(\frac{1}{\lambda} - \frac{1}{\lambda_0}\right) \right] d\lambda \quad \text{(S4)}$$

where $\lambda_0$ is the operating center wavelength and $P_0$ is an arbitrary power. Likewise, $P_{min}$ can be calculated with Eq. S4, except with a $\sin^2$. **Figure S6(B)** shows how for bandwidths > 0.7 GHz our experiment visibility results converge to a Gaussian numeric model (with 13.7 dB noise), but below this bandwidth our visibility deviates and matches closer to a Lorentzian numeric model (with 17.5 dB noise). Note that the data presented in **Fig. S6(B)** is different than the data presented in **Fig. 2(E)**. Noise is included to realistically model our photodetector measurements, as Eq. S3 would determine $P_{min} \approx 0$ for a very narrowband linewidth, which is clearly unphysical.

Figure 2(D) and **Figure S6(B)** both show broadening results up to ~2 GHz (~7 pm), which results in, at most, a drop of ~15% visibility. Theoretically, the ORD wavemeter can measure spectral broadening up until $V \approx 0$, which indicates that there is a limit to the total spectral broadening the ORD wavemeter can detect. However, we can overcome this limit and extend the total bandwidth detection. By shortening fiber length and/or using lower order modes the sinusoidal power distribution in Eq. S4 widens, spreading over a larger span of wavelengths. This would allow us to measure larger changes in $\delta\lambda$ for similar drops in visibility, hence extending the range over which we can broaden our spectrum.

5) <u>Mechanical and Thermal Perturbations</u>

We determine stability of our ORD mapping by assuring that for a single wavelength ($\lambda_0$) its corresponding polarization angle $\gamma_0$ does not drift over time. From Figure 3(A) we see that the ORD mapping is far more stable than the PM-fiber wavelength to polarization mapping[27]. We consider two cases, mechanical and thermal perturbations, where we used a layout identical to that presented in the Methods section, with the only exceptions being that we laid out our 40 m of vortex fiber on a stirring hotplate (Thermolyne Cimarec 2) and used two Ge Thorlabs S122C photodetectors sampling power (hence $\gamma$). For the mechanical perturbations, we laid out the fiber on cardboard base and placed it on the Cimarec, and used the maximum stirring function to physically shake the fiber. Over the course of 5 minutes we measured the OA angle [see **Fig. 3(A)**] and determined that the ORD wavemeter remained stable. Next we introduced thermal perturbations to the vortex fiber, now placed on an aluminum base, with the Cimarec hotplate functionality. For this stability test, we slowly increased temperature of the base from 22 °C to 60 °C, heating the fiber, and measured OA angle as a function of temperature [see **Fig. 3(B)**]. We measured temperature change in real-time using a thermistor connected to a Keysight 34461A Digital Multimeter. Unlike the previous stability test, we do see that OA angle changes as a clean periodic function with respect to temperature (i.e. time). This predicable and systematic behavior due to temperature fluctuations indicates that with careful calibrations, this thermic effect could be accounted for and mitigated.

6) <u>SOI $\Delta n_g$ modal dependence</u>

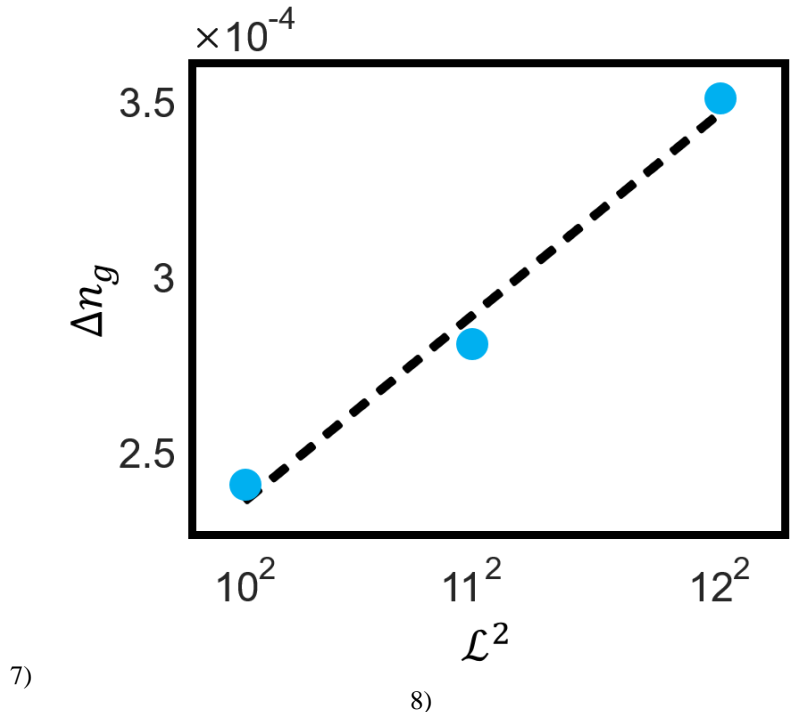

7)

8)

9) **Fig. S7.** Experimental validation that the group index difference between SOa and SOaa modes $\Delta n_g$ in vortex fiber follows the $\mathcal{L}^2$ mode dependence as theoretically predicted by SOI index splitting.

The spin-orbit interaction splitting of effective index between SOa and SOaa vortex fiber modes, $\Delta n_{eff}$, has an OAM modal dependence proportional to $\mathcal{L}^2$ in these vortex fibers[24] [see Eq. 2], as determined by perturbative theory[12]. For non-dispersive media we can also approximate that the difference in group index between these modes, $\Delta n_g$, should have the same dependence. We can numerically calculate $\Delta n_g$ using a finite-difference waveguide solver on the vortex fiber index profile[50], which is measured using an interferometry-based fiber profiler (Interfiber Analysis IFA-100) at 633nm. However, these simulations may be prone to measurement uncertainties, and for this reason we require experimental verification. Since $\Delta n_g$ is a key factor in establishing ORD strength [see Eq. 5], we can calculate it directly from our $\alpha$ calibration factor. By preforming separate calibrations [as discussed in the Methods section] for multiple $\mathcal{L}$-OAM modes, we can compare how $\Delta n_g$ changes with topological charge. **Figure S7** shows results for OAM modes $\mathcal{L} = 10$, 11, and 12, demonstrating excellent agreement of experiment with theory, as the derived $\Delta n_g$ values from ORD calibration is clearly proportional to $\mathcal{L}^2$.

10) <u>Interference of Gaussian and OAM</u>

In **Figure 4(B)** we illustratively depict the angular momentum through the use of a spiral interference between an OAM and Gaussian beam. The spiral arms, or parastiches, reveal the number of $2\pi$ phase rotations comprising the OAM beam's helical phase front, equal to its topological charge $\mathcal{L}$. These spiral patterns were created by reflecting a diagonally polarized beam off a SLM. The beam's $\hat{x}$-polarized projection is converted to an OAM beam while the $\hat{y}$-polarized projection remains a Gaussian beam. Passing these beams through a 45° linear polarizer recombines them to generate the spiral interference pattern. The spirals were then imaged on a camera to capture its intensity profile [see **Fig. S8**].

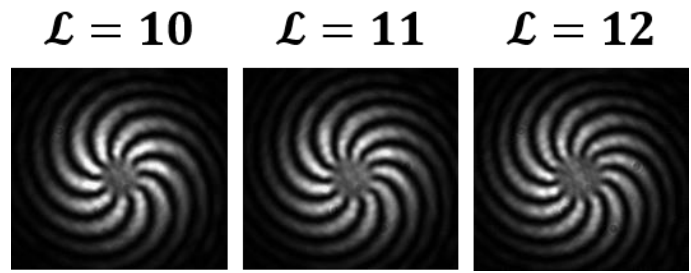

$$\boldsymbol{\mathcal{L} = 10} \qquad \boldsymbol{\mathcal{L} = 11} \qquad \boldsymbol{\mathcal{L} = 12}$$

**Fig. S8.** The intensity profiles of spirals created by interfering $\mathcal{L}$ = 10, 11, and 12 OAM beams with a Gaussian beam. The spirals highlight the angular momentum carried by the helical phase front of OAM beams.